\documentclass[prb,aps,twocolumn,showpacs]{revtex4}

\usepackage{graphicx}
\usepackage{dcolumn}
\usepackage{bm}
\usepackage{amsfonts}

\begin{document}

\title{Anomalous Magnetoresistance in
Pb-doped Bi$_2$Sr$_2$Co$_2$O$_y$ Single Crystals}
\author{X. G. Luo, X. H. Chen$^{\ast}$, G. Y. Wang, C. H. Wang,
X. Li, W. J. Miao, G. Wu, and Y. M. Xiong} \affiliation{Hefei
National Laboratory for Physical Science at Microscale and
Department of Physics, University of Science and Technology of
China, Hefei, Anhui 230026, People's Republic of China\\}
\date{\today}

\begin{abstract}
Magnetoresistance (MR) of the Bi$_{2-x}$Pb$_x$Sr$_2$Co$_2$O$_y$
($x$=0, 0.3, 0.4) single crystals is investigated systematically.
A nonmonotonic variation of the isothermal in-plane and
out-of-plane MR with the field is observed. The out-of-plane MR is
positive in high temperatures and increases with decreasing $T$,
and exhibits a pronounced hump, and changes the sign from positive
 to negative at a centain temperature. These results strongly suggest that
 the observed MR consists of two contributions: one \emph{negative} and one \emph{positive}
component. The isothermal MR in high magnetic fields follows a
$H^2$ law. While the negative contribution comes  from spin
scattering of carriers by localized-magnetic-moments based on the
Khosla-Fischer model.
\end{abstract}

 \maketitle

\section{INTRODUCTION}\vspace*{-2mm}
The triangular cobalt oxides attracted a great deal of interest
for the large thermoelectric power (TP) with low resistivity and
low thermal conductivity (thus the large thermoelectric figure of
merit $ZT=S^2T/\rho\kappa$) for the application reasons. A lot of
efforts has focused on the enhancement of the figure of
merit.\cite{Masset, Xu,Pelloquin} One important aspect for such
effort is to make out why the metallic oxides with triangular
CoO$_2$ layers have such unusually large TP comparing to the
conventional metal. Therefore, the work on understanding the
fundamental properties of this systems becomes especially
significant.

A number of results has been obtained on the transport and
magnetic properties of the triangular cobaltites, such as
Curie-Weiss susceptibility and temperature-dependent Hall
coefficient and anomalous magnetoresistance.\cite{Masset,
Yamamoto, Maignan, Hebert, Hervieu, Motohashi} Recently,
superconductivity was found in one of the promising thermoelectric
triangular cobaltite Na$_x$CoO$_2$ with $x$=0.35 by intercalating
water molecules into between the Na$^+$ and CoO$_2$
layers.\cite{Takada} Later, Foo et al. \cite{Foo} observed an
insulating resistivity below 50 K in the composition of $x$=0.5,
which is considered to be related to the strong coupling of the
holes and the long-range ordered Na$^+$ ions. The strong magnetic
field dependence of TP in Na$_x$CoO$_2$ provides an unambiguous
evidence of strong electron-electron correlation in the
thermoelectric cobalt oxides.\cite{Wang} The large TP with
metallic resistivity, superconductivity, charge ordering existing
with various $x$, displays a complicated and profuse electronic
state in Na$_x$CoO$_2$. This has inspired numerous theoretical and
experimental studies on the triangular cobalt oxides.

In this paper, we present new results on the magnetoresistance
(MR) of Bi$_{2-x}$Pb$_{x}$Sr$_{2}$Co$_{2}$O$_{y}$ ($x$=0.0-0.4)
((Bi,Pb)-Sr-Co-O) single crystals. It has been reported that there
exhibits large \emph{negative} MR in Ca$_{3}$Co$_{4}$O$_{9}$,
Bi$_{2-x}$Pb$_{x}$Sr$_{2}$Co$_{2}$O$_{y}$ and
[Bi$_{1.7}$Co$_{0.3}$Ca$_{2}$O$_{4}$]$^{\rm RS}$[CoO$_2$]$_{1.67}$
(Bi-Ca-Co-O),
[Pb$_{0.7}$A$_{0.4}$Sr$_{1.9}$O$_{1.9}$O$_{3}$]$^{\rm
RS}$[CoO$_{2}$]$_{1.8}$ (A=Hg,Co) \cite{Masset, Luo, Maignan,
Yamamoto, Pelloquin} while \emph{positive} MR in
Tl$_{0.4}$[Sr$_{0.9}$O]$_{1.12}$CoO$_{2}$,
[Bi$_{2}$Ba$_{1.8}$Co$_{0.2}$O$_{4}$]$^{\rm RS}$[CoO$_{2}$]$_{2}$
(Bi-Ba-Co-O) and Na$_{0.75}$CoO$_2$.\cite{Hebert,
Hervieu,Motohashi} Very recently, a magnetic-field-induced
insulator-to-metal transition was observed by us in
oxygen-annealed Ca$_{3}$Co$_{4}$O$_{9}$. \cite{Luo} Large negative
MR is ascribed to suppression of spin
scattering.\cite{Masset,Yamamoto,Maignan,Pelloquin}  The positive
$H^2$-dependence of MR in Na$_{0.75}$CoO$_{2}$ is attributed to
conventional orbital motion of carriers, \cite{Motohashi} while
linear positive MR is interpreted in an opening of
pseudogap.\cite{Hervieu} Up to now, the anomalous MR is not well
understood entirely yet. Further detailed investigation is
required to describe the physical nature of these triangular
cobaltites. For the easy control of doping level, Pb-doped
Bi$_2$Sr$_2$Co$_2$O$_y$ single crystals are chosen to be studied
systemically. The Pb-free sample, with the exact formula as
[Bi$_{0.86}$SrO$_2$]$_{2}^{\rm
RS}$[CoO$_2$]$_{1.82}$,\cite{Leligny} is paramagnetic down to 2
K.\cite{Yamamoto,Sugiyama} With doping by Pb, weak ferromagnetism
is induced, which is considered to originate from a
canted-antiferrimagnetism or to coexist with
spin-glass.\cite{Yamamoto,Yamamoto1} The coexistence of
ferromagnetic and antiferromagnetic interaction makes this system
intriguing. In this paper, the magnetoresistance, measured in
in-plane and out-of-plane configurations with the fields up to
13.5 T, is observed to exhibit nonmonotonic field-dependent
behavior. The MR shows a complicated dependence of the field and
temperature. Its magnitude is enhanced initially with increasing
$H$, and reaches a maximum at a certain field. The $T$ dependence
of the out-of-plane MR is positive in high temeratures, and shows
a pronounced hump with decreasing $T$. These results suggest the
MR comes from the contribution of negative and positive
components, which is strongly dependent of temperature and field.

\section{EXPERIMENT}
\vspace*{-2mm} The Bi$_{2-x}$Pb$_x$Sr$_2$Co$_2$O$_y$ ($x$=0.0,
0.3, 0.4) single crystals were grown by a self-flux method.
High-purity Bi$_2$O$_3$, PbO, SrCO$_3$ and Co$_3$O$_4$ were mixed
with a nominal mole ratio of Bi:Pb:Sr:Co=2-$x$:$x$:2:2 ($x$=0.5
and 0.6) and preheated at 800 $^\circ$C for 24 h. Additional
Bi$_2$O$_3$ as the flux was mixed with the obtained precursors
carefully. The mixture was melt at 1050 $^\circ$C for 2 h, and
followed a slow cooling procedure with a cooling rate of 4-5
$^\circ$C/h to 800 $^\circ$C, then cooled by furnace. Platelike
crystals were easily cleaved from the melt. The actual composition
of 0.3 and 0.4 was determined by inductively coupled plasmas (ICP)
atomic emission spectroscopy for the nominal composition of x=0.5
and 0.6, respectively. The measurements of in-plane and
out-of-plane resistivity with a configuration described in ref. 16
were performed using the standard ac four-probe method. The
magnetic field was supplied by a superconducting magnet system
(Oxford Instruments). \vspace*{-2mm}

\section{EXPERIMENTAL RESULTS and DISCUSSION}
\vspace*{-2mm}
\subsection{Transport properties}
\vspace*{-2mm}
\begin{figure}[htbp]
\centering
\includegraphics[width=7.5cm]{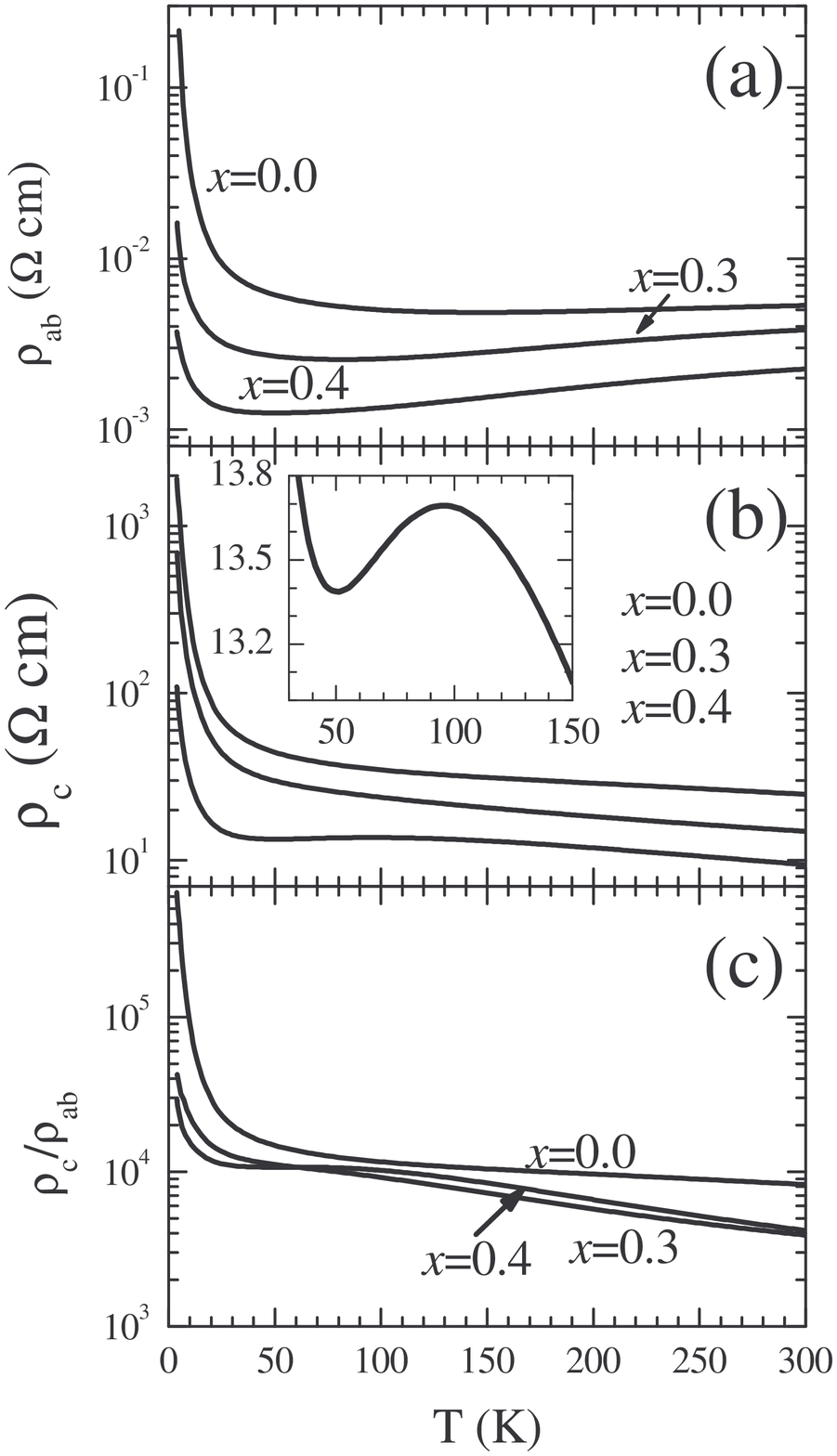}
\caption{\label{fig:epsart} The temperature dependence of in-plane
(a), out-of-plane (b) resistivity and the anisotropy
$\rho_{c}/\rho_{ab}$ (c).}\vspace*{-2mm}
\end{figure}

Figure 1(a) shows the temperature dependence of the in-plane
resistivity ($\rho_{ab}(T)$) of samples $x$=0.0, 0.3, and 0.4. The
crystals show metallic behavior in high temperatures.
$\rho_{ab}(T)$ exhibits a minimum at 140 K for $x$=0.0, 70 K for
$x$=0.3, and 45 K for $x$=0.4, respectively. Below this
temperature the crystals show a diverging resistivity. The
resistivity and the temperature corresponding to the resistivity
minimum decrease with increasing the content of the Pb
substitution for Bi. The ratio $\rho_{ab}(T=4K)/\rho_{ab}(T=300K)$
decreases also with enhancing the doping level of Pb. These
indicate that the Pb doping induces holes into the system. These
results are consistent with the previous report.\cite{Yamamoto}
Figure 1(b) shows the temperature dependence of the out-of-plane
resistivity ($\rho_{c}$(T)) of the samples. For the samples with
$x$=0.0 and $x$= 0.3, $\rho_{c}(T)$ shows insulator-like behavior
in the whole temperature range. It increases slightly above about
50 K and enhances suddenly below 50 K with decreasing temperature.
For the sample with $x$=0.4, $\rho_{c}(T)$ shows an insulator-like
behavior above 100 K and below 100 K it shows a metallic behavior
(d$\rho_{c}$/dT$>$0). With further decreasing temperature down to
30 K, $\rho_{c}(T)$ shows a reentrant insulating behavior and
increases sharply with decreasing temperature. It shows a broad
maximum at $T_{M}\approx$ 100 K. Such a broad maximum is very
similar to that observed in
(Bi$_{0.5}$Pb$_{0.5}$)$_{2}$Ba$_{3}$Co$_{2}$O$_{y}$ ($T_{M}
\approx$ 200 K) and NaCoO$_{2}$ ($T_{M}$ $\approx$ 180 K), where
it is thought to be an incoherent-coherent resistivity
transition.\cite{Valla} This transition was considered as a
crossover in the number of effective dimension from two to
three.\cite{Valla} The diverging resistivity in the low
temperatures has been attributed to the decrease of the effective
carrier number $n$ due to a pseudogap formation below 30-50
K.\cite{Itoh} In addition, the Hall coefficient was reported to
exhibit a sudden enhancement,\cite{Yamamoto, Yamamoto2} suggesting
a reduction of $n$ in low temperatures. Another point of view is
that the resistivity upturn and the sudden enhancement of Hall
coefficient below 50 K are associated with the magnetic ordering
in low $T$.\cite{Hebert}

Figure 1(c) shows the temperature dependence of the anisotropy
$\rho_{c}/\rho_{ab}$ for the three samples. The three samples show
close values of the anisotropy. The anisotropy shows a weak
temperature dependence above 50 K for the samples.  While the
anisotropy increases sharply below 30 K. It is addressed that the
anisotropy for the sample $x$=0.4 saturates below 100 K, which
coincides with the "incoherent-coherent" transition temperature.
The temperature-independent anisotropy between 100 and 30 K and
the peak in $\rho_c(T)$ for x=0.4 sample give evidence for
existence of the dimensional crossover from two to three. In order
to make clear the physics of the diverging resistivity and
"incoherent-coherent" transition, it requires angle-resolved
photoemission to determine the electronic structure.

\begin{figure*}[t]
\begin{minipage}[c]{0.48\textwidth}
\centering
\includegraphics[width=0.75\textwidth]{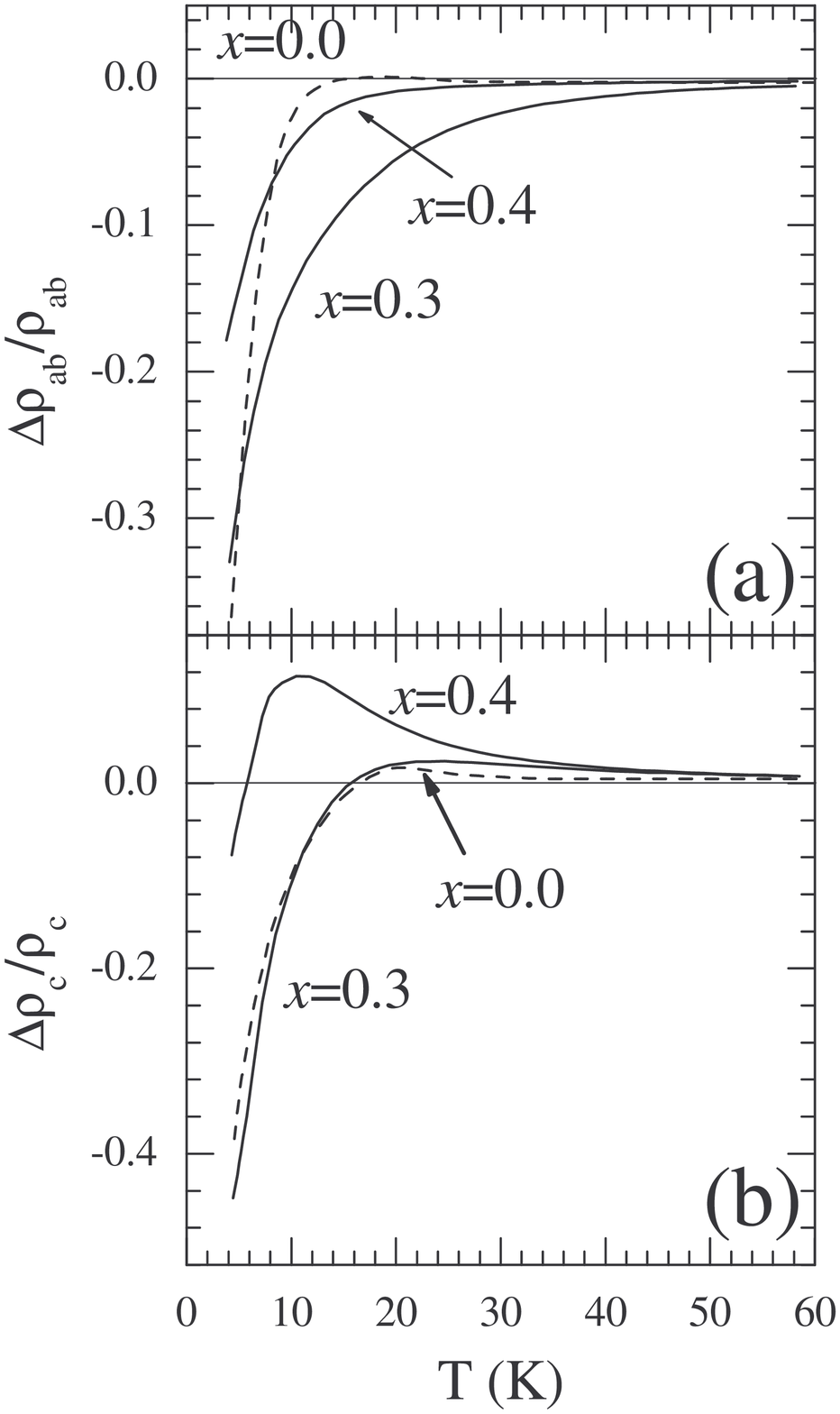}
\caption{The in-plane (a) and out-of-plane (b) magnetoresistance
as a function of temperature at 13.5 T for the two samples. The
field is applied along c-axis.} \label{fig2}
\end{minipage}
\hspace{0.02\textwidth}
\begin{minipage}[c]{0.48\textwidth}
\centering
\includegraphics[width=0.75\textwidth]{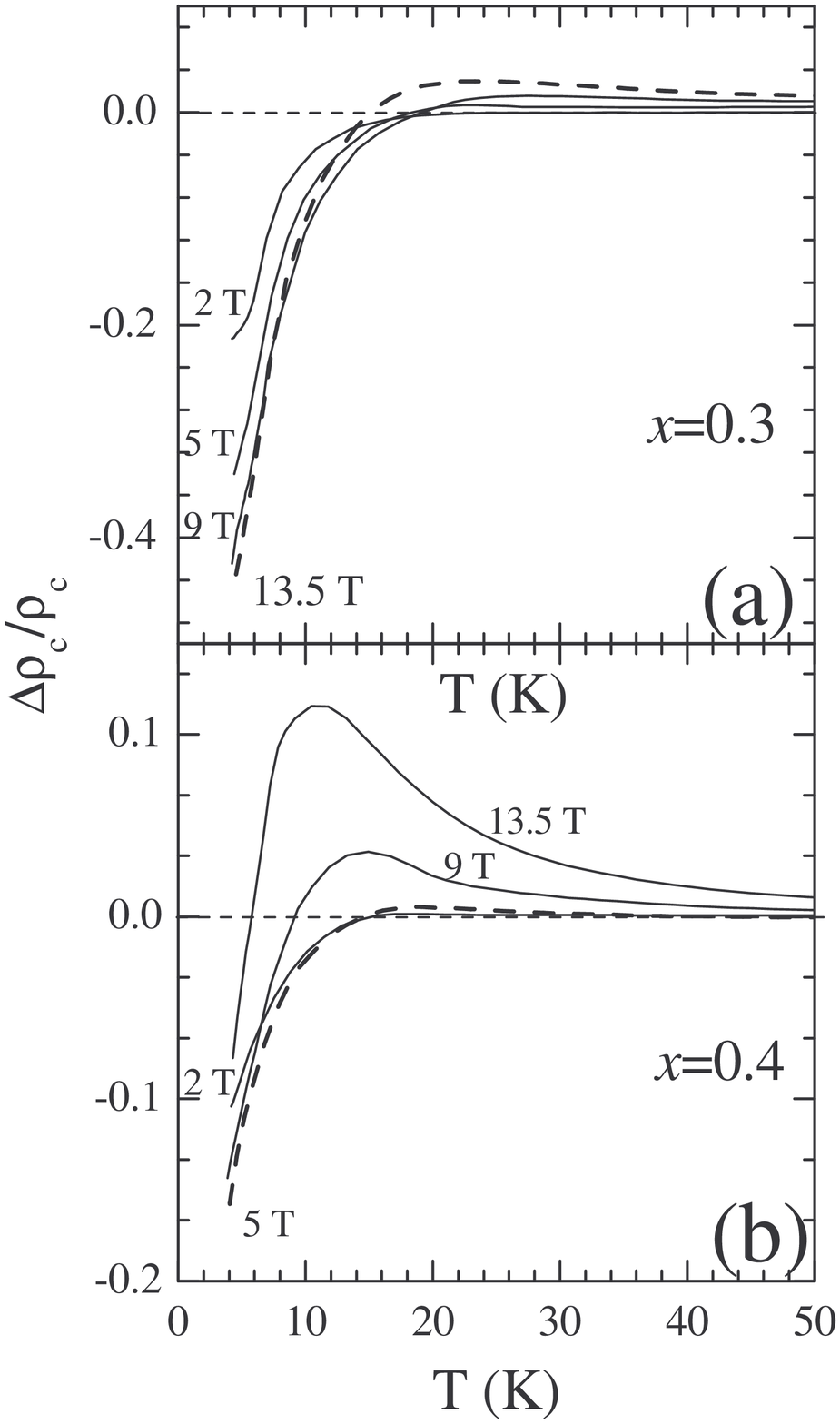}
\caption{The out-of-plane magnetoresistance as a function of
temperature at various magnetic fields for the sample $x$=0.3 (a)
$x$=0.4 (b). The field is applied along c-axis.} \label{fig3}
\end{minipage}
\end{figure*}

\vspace*{-2mm}
\subsection{Magnetoresistance}

Figure 2 shows the evolution of $\Delta\rho_{ab}/\rho_{ab}$ and
$\Delta\rho_{c}/\rho_{c}$
($\Delta\rho/\rho=((\rho(H)-\rho(0))/\rho(0)$) with varying
temperature at the field of 13.5 T. Above 12 K, the in-plane MR
for $x$=0.0 is very small and negative, and its magnitude
increases sharply below 12 K with decreasing temperature. The
negative MR reaches 37$\%$ at 4 K. The in-plane MR is negative and
its magnitude monotonously increases with decreasing temperature,
and reaches 33$\%$, and 18$\%$ at 4 K for 0.3 and 0.4 sample,
respectively. However, the out-of-plane MR shows anomalous
features compared to the in-plane MR. The out-of-plane MR is
positive in high temperatures for the samples. The MR first
increases with decreasing temperature, and exhibits a broad hump.
With further decreasing temperature, the MR changes the sign from
positive to negative, and its magnitude begins to increase
sharply.

In order to understand the anomalous behavior in out-of-plane MR,
it was systematically studied in the different magnetic fields.
The out-of-plane MR for the samples with $x$=0.3 and 0.4 as a
function of temperature at various magnetic fields is shown in
Fig. 3. The MR is positive in high temperatures as observed in
Fig.2, and increases monotonically with increasing magnetic field.
Broad humps of the positive MR can be observed at all magnetic
fields for the two samples. The position of the humps shifts to
lower temperature with enhancing magnetic field. With further
decreasing $T$, the MR becomes negative in low temperatures. The
temperature for the MR passing through zero decreases monotonously
with increasing the field. The negative MR exhibits complex
behavior at various fields. The magnitude of MR for $x$=0.3 varies
with field monotonically at 4 K, while exhibits maximum at 9 T as
$T$ is in the temperature range from 8 K to 18 K. The magnitude of
MR for the $x$=0.4 reaches maximum around 5 T at 4 K. These
results suggests that the observed MR comes from two
contributions, i.e. one negative and one positive contribution to
the total MR. In high temperatures, the positive component is
predominant, while the negative component grows more rapidly than
the positive one with decreasing $T$ and MR has a negative value
in low temperatures.

\begin{figure*}[htbp]
\begin{minipage}[c]{0.48\textwidth}
\centering
\includegraphics[width=0.75\textwidth]{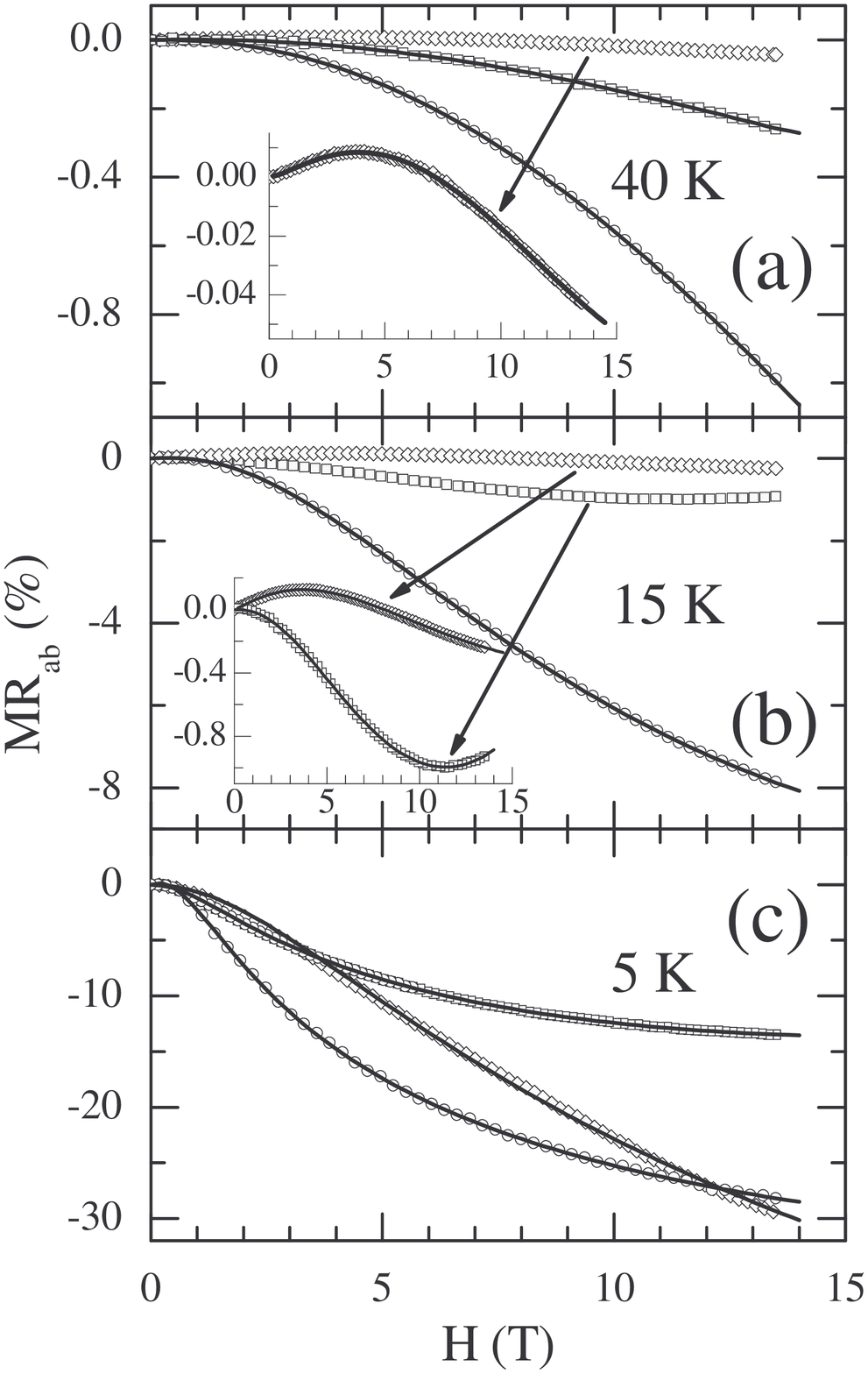}
\caption{The isothermal in-plane magnetoresistance as function of
magnetic field at 40 K, 15 K, and 5 K for the samples with {\it
x}=0.0 ($\diamond$), {\it x}=0.3 ($\circ$) and {\it x}=0.4 ({\tiny
$\square$}), where MR=[$\rho(H)-\rho(0)$]/$\rho(0)\times100\%$,
the subindex ab is referred to the in-plane case. The solid lines
are the data fitted by using Eq. (1). The field is applied along
c-axis.} \label{fig5}
\end{minipage}
\hspace{0.02\textwidth}
\begin{minipage}[c]{0.48\textwidth}
\centering
\includegraphics[width=0.75\textwidth]{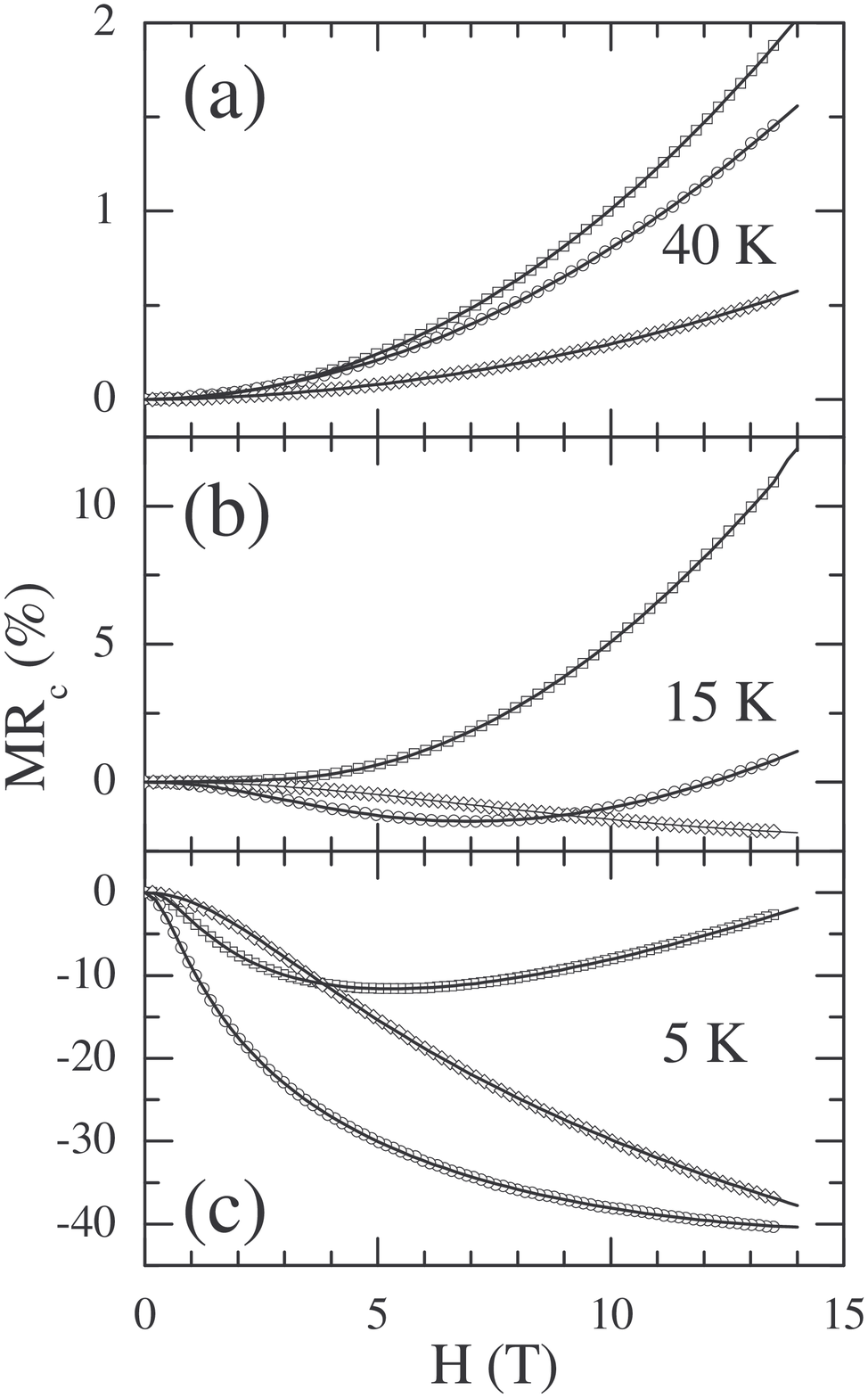}
\caption{The isothermal out-of-plane magnetoresistance as function
of magnetic field at 40 K, 15 K, and 5 K for the samples with {\it
x}=0.0 ($\diamond$), {\it x}=0.3 ($\circ$) and {\it x}=0.4 ({\tiny
$\square$}). The subindex c of MR is referred to the out-of-plane
configuration. The solid lines represent the data fitted by Eq.
(1). The field is applied along c-axis.}\label{fig6}
\end{minipage}
\end{figure*}

In order to clearly understand the complicated behavior shown in
Fig.3, isothermal MR is studied by sweeping the fields up to 13.5
T at different temperatures. Figure 4 shows the isothermal
in-plane MR at 5, 15 and 40 K for the samples with $x$=0.0, 0.3,
and 0.4. There are some salient features of the MR. (I) The MR at
40 K and 15 K for the crystal with $x$=0.0 first increases with
increasing field, and reaches a maximum at 4 T (for both of the
temperatures), then decreases monotonically. Such a MR has not
been observed in the misfit-layered cobaltites. (II) All the other
MR curves are negative, with the magnitude increasing with
decreasing temperature. The magnitude of these MR increases
monotonically with increasing magnetic field except that at 15 K
in the sample $x$=0.4. (III) The magnitude of the MR at 15 K for
$x$=0.4 first increases with increasing magnetic field, and
reaches a maximum ($\sim$10$\%$) at 11.4 T, then decreases with
further increasing field (see the inset in Fig. 4(b)). It suggests
the presence of a positive contribution in additional to the
negative MR in this case. Such an anomalous MR has not been
observed in triangular cobaltites previously. Only large monotonic
negative MR has been found in (Bi,Pb)-Sr-Co-O, Bi-Ca-Co-O up to 9
T \cite{Maignan, Yamamoto} and Ca$_3$Co$_4$O$_9$ up to 14 T.
\cite{Luo} (IV) The crystal with the $x$=0.3 have the largest
negative MR for all temperature except that above 12 T at 5 K.
This is consistent with the data shown in Fig.2(a), where an
intersection is observed at about 5.1 K of $x$=0.0 and $x$=0.3 MR
curves. According to the susceptibility and $\mu$SR
measurements,\cite{Yamamoto,Sugiyama} the Pb-free crystal is
almost paramagnetic down to 2 K. Ferromagnetism would be induced
by Pb doping and the transition temperature ($T_c$) increases with
increasing Pb doping level, with $T_c$=3.2 K for $x$=0.44 and 5 K
for $x$=0.51, respectively. Thus in the present samples, the $T_c$
is less than 3.2 K and beyond our measurement temperature range.
The relative large magnitude of the negative MR at 15 K and 5 K
may suggest the presence of short-range ferromagnetic correlation
far above $T_c$.

\begin{table*}[htbp]
\caption{\label{tab:table1}Values of the parameters to fit MR with
Eq. (1). The out-of-plane MR for $x$=0.4 at 40 K can fitted only
using the second term in Eq. (1). }
\begin{ruledtabular}
\begin{tabular}{llllll}
&$T(K)$ & $A_{1}$ & $A_{2}$  &$B_{1}$ & $n$\\ \hline\ $x$=0.0
in-plane & 5 & 3.51 & 0.23216 & 0& 0 \\ \
&15 & 1.41197 & 0.09207 & 0.26845 & 1.19685 \\\
&40 &1.00262 & 0.05016 & 0.07236 & 1.59962 \\ \\
$x$=0.0 out-of-plane & 5 & 3.57139 & 0.3064& 0 & 0 \\\
& 15 & 1.94482 & 0.10354 & 0.17494 & 1.66214 \\\
&40 &0.61187 & 0.06884 & 0.07823 & 1.85638 \\ \\
$x$=0.3 in-plane &5 &2.84001 &0.73358&1.09044 & 0.77283 \\\
&15 &2.78349 &0.1563
&0.39737 &1.34744 \\\
 &40 &2.59917 &0.03643 &0.07917 &1.65783\\ \\
 $x$=0.3 out-of-plane &5   &2.78462 &1.48822 &0.29458 &1.64505\\\
 &15  &1.74818 &0.23088 &0.28285 &1.77112\\\
 &40 &0.42274 &0.16466 &0.13152 &1.77808\\ \\
$x$=0.4 in-plane  &5 &1.9198 &0.62776 &0.08762 &2.19737\\\
 &15  &1.6376 &0.12996 &0.17557 &1.73868\\\
  &40 &1.04661 &0.049025 & 0.05667 &1.45842\\ \\
 $x$=0.4  out-of-plane &5 &2.50353 &1.09468 &1.24977 &1.14875\\\
 &15 &1.65991 &0.22107 &0.3908 &1.81456\\\
 &40 &0 &0 &0.09348 &2.06279\\
\end{tabular}
\vspace*{-2mm}
\end{ruledtabular}
\end{table*}
\vspace*{-2mm}

In Fig. 4(b), the nonmonotonic field-dependence of MR is anomalous
for triangular cobaltites, which has not been observed previously
in these systems. Such a nonmonotonic MR is more obvious in the
out-of-plane MR. Figure 5 shows the evolution of the isothermal
out-of-plane MR with magnetic field along c-axis at 5, 15, and 40
K for the three samples. The MR is positive at 40 K for all the
samples (in contrast to the in-plane MR), and increases with
increasing Pb doping level. The MR at 15 K is positive for
$x$=0.4, while negative for the Pb-free crystal. These two samples
exhibits monotonic MR at 15 K. On the contrary, the MR for $x$=0.3
at 15 K is first negative and its magnitude increases with
increasing magnetic field, and reaches a maximum ($\sim$1.43$\%$)
at about 6.9 T, then decreases and passes through zero at 12.5T.
The MR at 5 K is negative for all the crystals. The MR for $x$=0.0
and 0.3 decreases monotonically with increasing $H$, while the MR
of $x$=0.4 exhibits an analogous behavior to that of $x$=0.3 at 15
K, the magnitude of which first increases with enhancing $H$, and
reaches a maximum ($\sim$11.6$\%$) around 5.3 T, then decreases.
The above results are well consistent with the evolution of
magnetoresistivity with $T$ in Fig. 3. It should be pointed out
that for the sample $x$=0.4 the positive MR at 15 K is larger than
that at 40 K, which is consistent with the hump observed in Fig.
3. The results above strongly suggest \emph{the presence of a
positive contribution in MR}.

Therefore, an expression consisting of negative and positive
contributions for MR could be used to describe the anomalous
nonmonomonic MR. It is found that all the isothermal MR can be
well fitted by the following expression:
\begin{equation}
\frac{\Delta\rho}{\rho}=-A_1^2\ln(1+A_2^2H^2)+B_1^2H^n.
\end{equation}
where $A_1$, $A_2$, $B_1$, and $n$ are variable parameters for
fitting. The fitted data are plotted in Fig. 4 and Fig. 5 by solid
lines. All the isothermal MRs are well fitted. The fitting
parameters are listed in Table I. The fitting parameters $A_1$ and
$A_2$ increase with decreasing temperature. $A_2$ is very small at
40 K, indicating that the negative contribution of MR is slight.
For the out-of-plane MR of $x$=0.4, the first term is even zero.
The $A_1$ and $A_2$ shows systematical temperature dependence, but
no systematically concentration dependence is observed. Calculated
with these values of $A_1$ and $A_2$, the negative component of MR
is the largest in magnitude for $x$=0.3 at the three temperatures.
Compared to $A_1$ and $A_2$, $B_1$ and $n$ varies more complexly.
Neither systematical temperature dependence nor systematical
carrier concentration dependence is found. For $x$=0.0, this term
is absent at 5 K. At the other two temperatures, $B_1$ is larger
at 15 K and $n$ is larger at 40 K. For the other two crystals,
with decreasing $T$, $B_1$ increases, while $n$ decreases, except
in the in-plane case of $x$=0.4. For the in-plane MR of $x$=0.4,
with decreasing $T$, $B_1$ decreases, while $n$ increases. The
calculated results for this term is rather complex. In the
out-of-plane MR for $x$=0.3, $B_1$ decreases while $n$ increases
as the temperature is enhanced. While the calculated positive
component is the largest at 15 K. In the out-of-plane MR for
$x$=0.4, $B_1$ and $n$ have the same evolution with temperature as
that for $x$=0.4, while calculated positive component is the
largest at 5 K. The increase of $n$ enhances the dependence of the
positive component on the magnetic field, while the decrease of
$B_1$ reduces it: they have the opposite effect on the positive
component. This is the reason for the complex behavior of positive
component of MR with temperature and carrier concentration. In
general, the value of the positive component is relatively small
at 40 K, and increases as the temperature decreases. This is
consistent with the pronounced hump in the out-of-plane MR vs. T
curves.

\subsection{The possible origin of the anomalous MR}
\vspace*{-2mm}
\subsubsection{The negative component}
The First term in Eq. (1) comes from a semiempirical expression
proposed by Khosla and Fischer \cite{Khosla} and has been
previously used to explain the negative MR observed in $n$-type
CdS, \cite{Khosla} n-type Si \cite{Khosla1} and
(In,Mn)As.\cite{May} The basis for this formula is Toyozawa's
localized-magnetic-moment model of magnetoresistance, where
carriers in an impurity band are scattered by the localized spin
of impurity atoms.\cite{Toyozawa} It is derived from
third-perturbation expansion of the $s$-$d$ exchange Hamiltonian
in this local-magnetic-moment model of Toyazawa.
\cite{Khosla,Toyozawa} Well agreement of the MR with the Eq. (1)
reflects that the negative MR comes from the interaction between
conducting carriers and localized magnetic moments, and it reveals
a decrease of spin-dependent scattering of carriers in magnetic
field. This model requires a separation of conducting carriers and
localized magnetic moments. According to the photoemission and
x-ray-absorption spectroscopy measurements in the misfit-layered
(Bi,Pb)-Sr-Co-O,\cite{Mizokawa} both Co$^{3+}$ and Co$^{4+}$ have
low spin configuration. The electrons of Co$^{3+}$ and Co$^{4+}$
locate in the $t_{2g}$, and the three-folded $t_{2g}$ is split
into one $a_{1g}$ subband and two $e'_{g}$ subbands due to the
rhombohedral crystal field.\cite{Singh} In Ref. 25, it has been
pointed out that holes locate mainly in the $a_{1g}$ subband,
which are strongly coupled to the lattice and become localized
holes. A minority of holes locate in the $e'_g$ subbands. The
former are "heavier" than the latter. The local magnetic moments
are attributed to the $a_{1g}$ holes \cite{Yamamoto} due to the
strong electron-phonon coupling while $e'_{g}$ holes is conducting
carriers responsible for the relative low resistivity. Therefore,
It is inferred that the conducting carriers formed by the $e'_{g}$
holes interact with the localized magnetic moments from the
$a_{1g}$ holes through $s$-$d$ exchange. Actually, such anomalous
MR with nonmonotonic field dependence as shown in Fig. 5 has been
previously observed in R$_2$Ni$_2$Si$_5$ (R=Pr, Dy, Ho, and Er)
compounds, which was attributed to the presence of short-range
ferromagnetic order in the literatures.\cite{Mazumda1,Mazumda2}
However, in the Pb-free crystal, it is reported to be paramagnetic
down to 2 K.\cite{Yamamoto,Sugiyama} Therefore, it seems to be
difficult to be understood by the presence of short-range
ferromagnetic order. Using Khosla-Fischer model with $s$-$d$
exchange in localized-magnetic-moment model, the negative MR in
these crystals may be well understood.

\begin{figure}[htbp]
\centering
\includegraphics[width=0.4\textwidth]{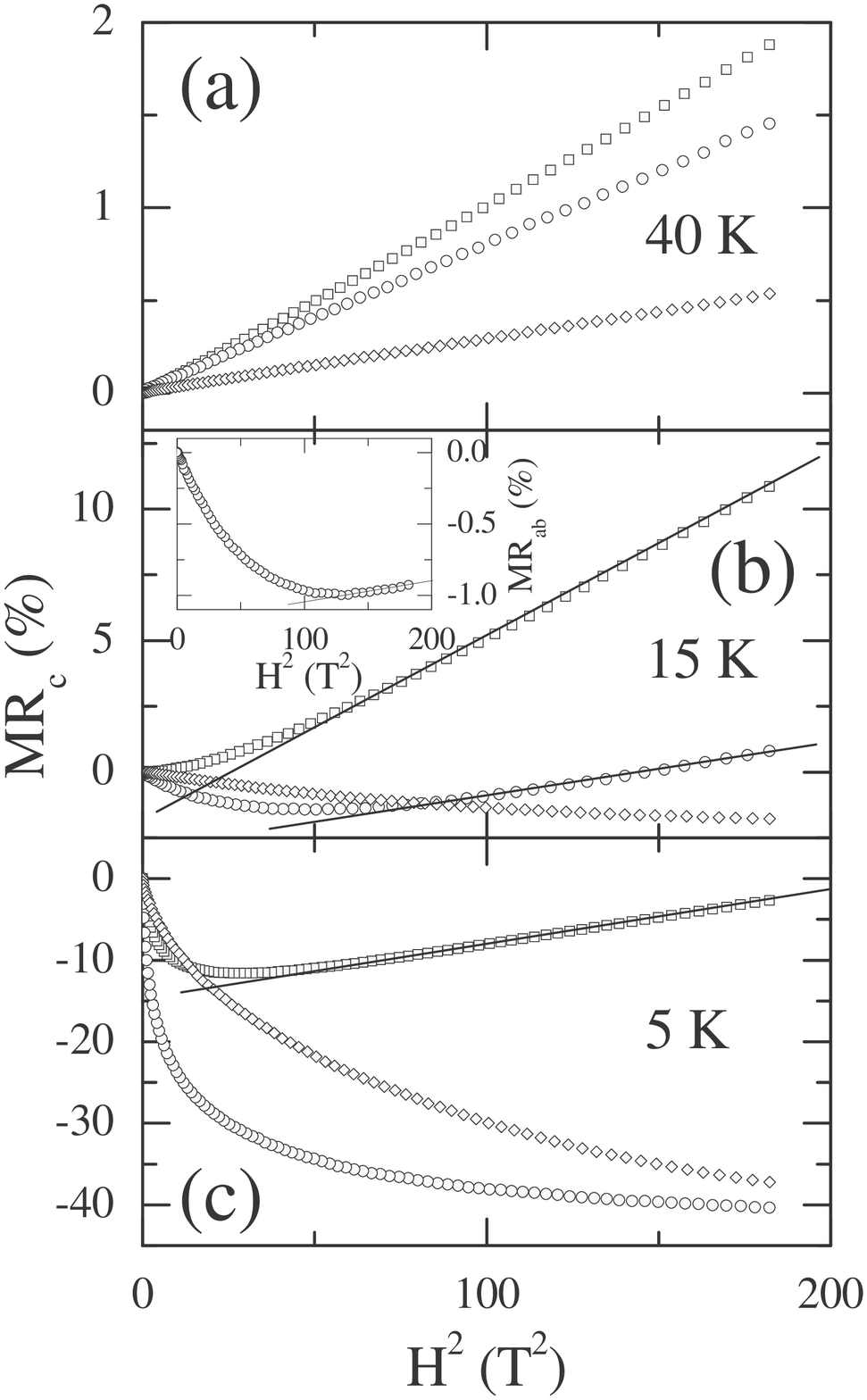}
\caption{The square magnetic dependence of the isothermal
out-of-plane MR at 5, 15, and 40 K for the crystals with {\it
x}=0.0 ($\diamond$), {\it x}=0.3 ($\circ$) and {\it x}=0.4 ({\tiny
$\square$}). The inset shows the in-plane MR at 15 K for {\it
x}=0.4.The field is applied along c-axis.}\label{fig6}
\end{figure}
\begin{figure}[htbp]
\centering
\includegraphics[angle=-90,width=0.48\textwidth]{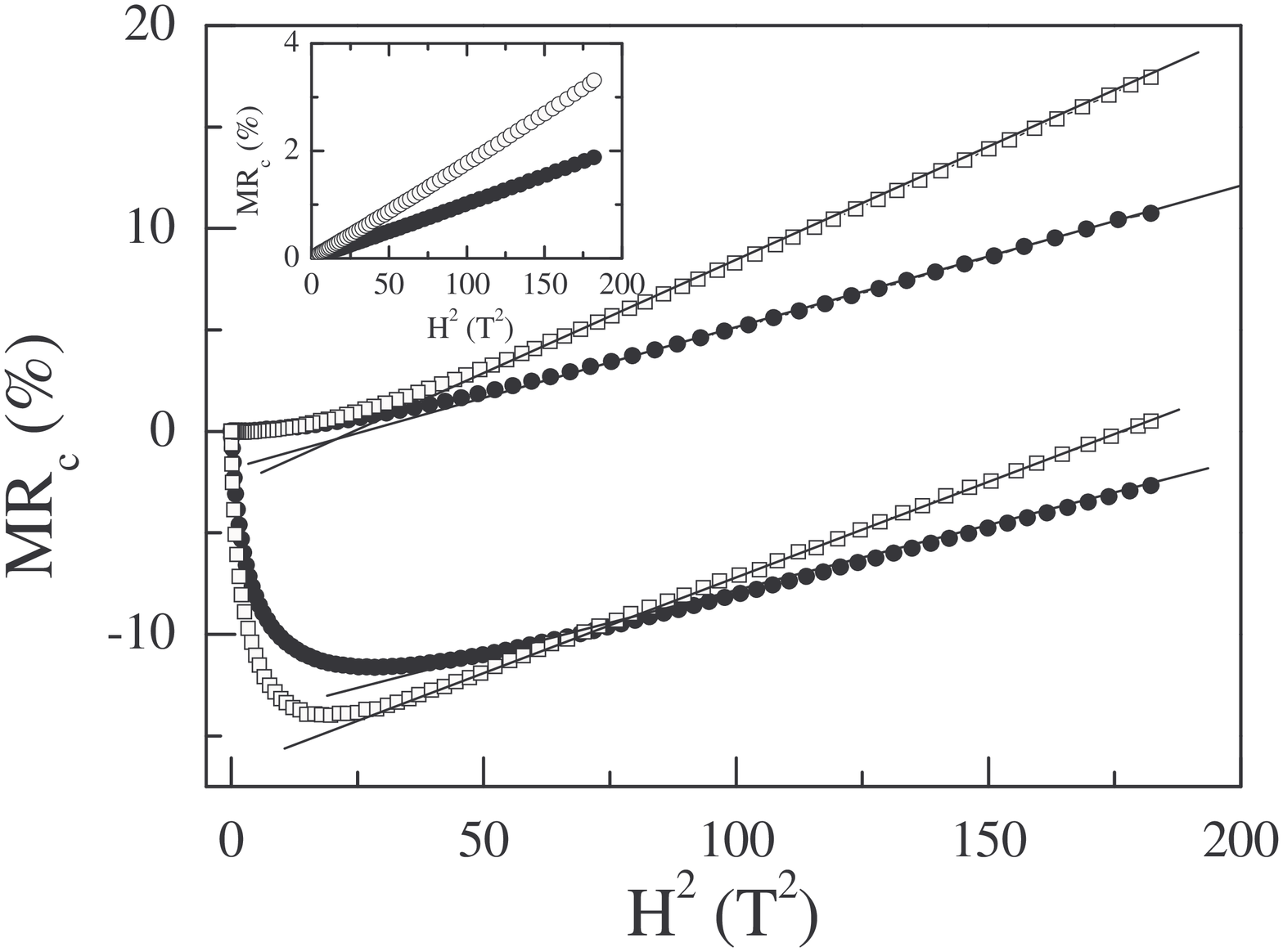}
\caption{The square magnetic field dependence of the isothermal
transverse ({\tiny $\square$}) and longitudinal ($\bullet$)
out-of-plane MR at 5, 15, and 40 K (the inset) for the crystal
with {\it x}=0.4. Transverse and longitudinal configurations
correspond to the field applied with ab-plane and along c-axis,
respectively.} \label{fig7} \vspace*{-5mm}
\end{figure}

\subsubsection{Positive component}
The hump feature in the out-of-plane MR vs $T$ curves and the
nonmonotonic isothermal MR strongly suggest the presence of the
positive contribution in addition to the negative component. The
second term in Eq. (1) gives the positive component of MR, which
is proportional to $H^n$, with $n$ varying from 0.77 to 2.19. The
nonmonotonic MR found in R$_2$Ni$_2$Si$_5$ (R=Pr, Dy, Ho, and Er)
is almost linear to $H$ in high fields,\cite{Mazumda1,Mazumda2}
which is ascribed to the blocks structure.\cite{Mazumda1} While
power law dependence of the positive component with power exponent
dependent dramatically of $T$ and sample is observed. In order to
understood this behavior, the out-of-plane MR is plotted in Fig. 6
by $H^2$ scales. It is intriguing that the upturn MR in high
fields exhibits clear $H^2$ behavior. In the inset of Fig. 6(b),
the in-plane MR at 15 K for $x$=0.4 also shows almost $H^2$
behavior in high fields. It seems that in high fields the upturn
MR is actually proportional to $H^2$. Large positive MR has ever
been observed in Na$_{0.75}$CoO$_2$ and
[Bi$_{2}$Ba$_{1.8}$Co$_{0.2}$O$_{4}$]$^{\rm
RS}$[CoO$_{2}$]$_{2}$,\cite{Motohashi,Hervieu} and is proportional
to $H^2$ in the former, while nearly linear to $H$ in the latter.
The $H^2$ dependence of positive MR in Na$_{0.75}$CoO$_2$ was
attributed to the conventional orbital motion of
carriers.\cite{Motohashi} However, this mechanism seems not to be
plausible in the present samples. The increase of MR with
decreasing temperature in Na$_{0.75}$CoO$_2$ is accompanying with
a dramatic enhancement in carrier mobility. However, in the
present crystals, for example, for $x$=0.4 the $H^2$ behavior of
the out-of-plane MR increases with decreasing $T$ from 40 K to 15
K, while the semiconducting resistivity suggests a reduction of
mobility of carriers, and this is in contrast to the case in
Na$_{0.75}$CoO$_2$. Furthermore, in classic Lorentz-force concept,
the orbital motion of carriers would give no contribution to the
longitudinal MR for spherical Fermi surfaces. Therefore, it
suggests that detailed information of topology of Fermi surface is
important to understand the observed magnetoresistive behavior.
Fig. 7 shows the transverse and longitudinal out-of-plane MR
plotted against $H^2$. The $H^2$ dependence in high field can be
clearly observed for all the curves. At each temperature, the
transverse MR is larger than longitudinal one in high fields. It
is interesting that the slopes of MR in high fields at 15 K and 40
K are almost the same in transverse or longitudinal case. This
suggests that positive contribution saturates below 15 K. This
positive contribution in MR is not understood yet. It may be
associated with the complex magnetic structure in this system. Two
possible magnetic configurations have been proposed by Yamamoto et
al.,\cite{Yamamoto,Yamamoto1} (1) A canted antiferromagnetic spin
structure and (2) the coexistence of spin-glass and
ferromagnetism, to interpret the weak ferromagnetism in Pb-doped
crystals. In these two pictures, antiferromagnetic interaction is
necessary. A positive MR with a slope following a $H^2$ law can be
expected for antiferromagnetic ordering.\cite{Yamada} The presence
of AF interactions over the length of the mean free path would
lead to a significant positive MR. More microscopic information
and theoretical work are required to understand this anomalous
nonmonotonic MR.

\section{CONCLUSION}
\vspace*{-2mm} We have observed an anomalous nonmonotonic
field-dependent behavior of MR in (Bi,Pb)-Sr-Co-O single crystals.
The MR exhibits a positive hump in high temperatures, following a
negative behavior at low temperature, and the magnitude of the
negative MR in low temperatures exhibits a maximum with magnetic
field. These results strongly suggest that the MR comes from two
contribution: one negative and one positive component. The
\emph{negative} component is described by the third-perturbation
expansion of the $s$-$d$ exchange Hamiltonian in
localized-magnetic-moment model of Toyozawa. The \emph{positive}
contribution follows a $H^2$ law in high fields. The understanding
of this anomalous nonmonotonic MR requires further experimental
and theoretical work on the microscopic mechanisms.\vspace*{-2mm}

\section{ACKNOWLEDGEMENT}\vspace*{-2mm}
This work is supported by the grant from the Nature Science Foundation of
China and by the Ministry of Science and
Technology of China (Grant No. NKBRSF-G1999064601), the Knowledge
Innovation Project of Chinese Academy of Sciences.\\

\vspace*{5mm} $^{\ast}$ \emph{Electronic address:} chenxh@ustc.edu.cn


\begin{thebibliography}{99}
\bibitem{Masset}
A. C. Masset, C. Michel, A. Maignan, M. Hervieu, O. Toulemonde, F. Studer, B.
Raveau, and J. Hejtmanek, Phys. Rev. B {\bf 62}, 166 (2000).

\bibitem{Xu}
G. J. Xu, R. Funahashi, M. Shikano, I. Matsubara, and Y. Q. Zhou, Appl. Phys.
Lett. {\bf 80}, 3760 (2002).

\bibitem{Pelloquin}
D. Pelloquin, A. Maignan, S. Hebert, C. Martin, M. Hervieu, C. Michel, L. B.
Wang and B. Raveau, Chem. Mater. {\bf 14}, 3100 (2002).

\bibitem{Maignan}
A. Maignan, S. Hebert, M. Hervieu, C. Machel, D. Pelloquin and
D. Khomskii, J. Phys: Condens. Matter {\bf 15}, 2711 (2003).

\bibitem{Yamamoto}
T. Yamamoto, K. Uchinokura, and I. Tsukada, Phys. Rev. B {\bf 65}, 184434
(2002).

\bibitem{Hebert}
S. Hebert, S. Lambert, D. Pelloquin and A. Maignan, Phys. Rev. B {\bf 64},
172101 (2001).

\bibitem{Hervieu}
M. Hervieu, A. Maignan,C. Michel, V. Hardy, C. Creon, and B. Raveau, Phys.
Rev. B {\bf 67}, 045112 (2003).

\bibitem{Motohashi}
T. Motohashi, R. Ueda, E. Naujalis, T. Tojo, I. Terasaki, T. Atake, M.
Karppinen, and H. Yamauchi, Phys. Rev. B {\bf 67}, 064406 (2003).

\bibitem{Takada}
K. Takada, H. Sakurai, E. Takayama-Muromachi, F. Izumi, R. A. Dilanian,
and T. Sasaki, Nature {\bf 422}, 53 (2003).

\bibitem{Foo}
M. L. Foo, Y. Y. Wang, S. Watauchi, H. W. Zandbergen, T. He, R. J. Cava, and
N. P. Ong, Phys. Rev. Lett. {\bf 92}, 247001 (2004).

\bibitem{Wang}
Y. Y. Wang, N. S. Rogado, R. J. Cava, and N. P. Ong, Nature {\bf 423}, 425
(2003).

\bibitem{Luo}
X. G. Luo, X. H. Chen, G. Y. Wang, C. H. Wang, Y. M. Xiong, H. B.
Song, H. Li, and X. X. Lu, cond-mat/0412298.

\bibitem{Leligny}
H. Leligny, D. Grebille, O. Perez, A. C. Masset, M. Hervieu, C.
Michel, and B. Raveau, C. R. Sci. Paris IIc, Chim {\bf 2}, 409
(1999); Acta Crystallogr., Sect. B: Struct. Sci. {\bf 56}, 173
(2000).

\bibitem{Sugiyama}
J. Sugiyama, J. H. Brewer, E. J. Ansaldo, H. Itahara, T. Tani, M.
Mikami, Y. Mori, T. Sasaki, S. Hebert, and A. Maignan, Phys. Rev.
Lett. {\bf 92}, 17602 (2004).

\bibitem{Yamamoto1}
I. Tsukada, T. Yamamoto, M. Takagi, and T. Tsubone, Jpn. J. Appl.
Phys. {\bf 39}, 6658 (2000).

\bibitem{Wang1}
C. H. Wang, L. Huang, L. Wang, Y. Peng, X. G. Luo, Y. M. Xiong, and X. H.
Chen, Supercond. Sci. Technol. {\bf 17}, 469 (2004).

\bibitem{Valla}
T. Valla, P. D. Johnson, Z. Yusof, B. Wells, Q. Li, S. M.
Loureiro, R. J. Cava, M. Mikamik, Y. Morik, M. Yoshimurak, and T.
Sasakik,, Nature {\bf 417}, 627 (2002).

\bibitem{Itoh}
T. Itoh, and I. Terasaki, Jpn. J. Appl. Phys. {\bf 39}, 6658 (2000).

\bibitem{Yamamoto2}
T. Yamamoto, I. Tsukada, M. Takagi, T. Tsubone, and K. Uchinokura,
J. Magn. Magn. Mater. {\bf 226-230}, 2031 (2001).

\bibitem{Khosla}
R. P. Khosla, and J. R. Fischer, Phys. Rev. B {\bf 2}, 4084
(1970).

\bibitem{Khosla1}
R. P. Khosla, and J. R. Fischer, Phys. Rev. B {\bf 6}, 4073
(1972).

\bibitem{May}
S. J. May, A. J. Blattner, and B. W. Wessels, Phys. Rev. B {\bf
70}, 073303 (2004).

\bibitem{Toyozawa}
Y. Toyazawa, J. Phys. soc. Jpn. {\bf 17}, 986 (1962).

\bibitem{Mizokawa}
T. Mizokawa, L. H. Tjeng, P. G. Schultze, G. A. Sawatzky, N. B.
Brooks, I. Tsukada, T. Yamamoto and K. Uchinokura, Phys. Rev. B
{\bf 64}, 115104 (2001).

\bibitem{Mazumda1}
C. Mazumda, A. K. Nigam, R. Nagarajan, L. C. Gupta, C. Godart, G. Chandra,
and R. Vijayaraghavan, Phys. Rev. B {\bf}, 6069 (1996).

\bibitem{Mazumda2}
C. Mazumda, R. Nagarajan, A. K. Nigam, K. Ghosh, S. Ramakrishnan, L. C. Gupta,
and B. D. Padalia, Physica B {\bf 339}, 216 (2003).

\bibitem{Singh}
D. J. Singh, Phys. Rev. B {\bf 61}, 13397 (2000).

\bibitem{Yamada}
H. Yamada, and S. Takada, J. Phys. Soc. Jpn. {\bf
34}, 51 (1973).

\end{thebibliography}
\end{document}